# Jodrell Bank and the British Astronomical Association: The Making of a Generation of Scientists

**Jeremy Shears**

## Abstract

The discovery of radar echoes from meteor trains at Jodrell Bank in 1945 marked the beginning of a new era in meteor astronomy and laid the foundations for the subsequent development of radio astronomy. This paper examines the close relationship between Bernard Lovell's emerging research group at Jodrell Bank and the British Astronomical Association (BAA) during the formative years, 1946 to 1951. It explores the contributions of the individuals who connected these two communities. Particular attention is given to the lives and scientific careers of Michael Ovenden, Gerald Hawkins, John Clegg, John Davies, Victor Hughes and Mary Almond, all of whom were BAA members. Their later achievements demonstrate the remarkable talents of this generation of scientists, whose work extended across radio astronomy, planetary science, palaeomagnetism, archaeomagnetism and computer science.

## Introduction

The chance discovery by Bernard Lovell (Alfred Charles Bernard Lovell, 1913–2012) of radar echoes from meteor trains at Jodrell Bank in December 1945 heralded the beginning of a new era in meteor astronomy. This led to a deep collaboration between Lovell and John Philip Manning Prentice (1903-1981), Director of the British Astronomical Association (BAA) Meteor Section, whose knowledge of visual meteor observation proved essential to interpreting the discoveries. Between 1946 and 1951, Prentice arranged for members of his Section, including himself, to visit Jodrell to conduct visual observations alongside Lovell's radar team.

The first observing sessions involving Meteor Section observers took place during the Perseid meteor shower in July and August 1946 (1). Prentice himself was not present; instead, he sent two observers from his team. Michael Ovenden and Gerald Hawkins were only twenty and eighteen years old, respectively. Ovenden was a university student, and Hawkins had just left school and was waiting to go to university. Both were deeply interested in astronomy and each went on to pursue research degrees and distinguished professional careers in astronomy and related fields.

From 1946 onwards, increasingly strong links developed between Lovell's team and the BAA. For example, John Guy Porter (1900–1981), Director of the Computing Section, calculated meteor trajectories that proved invaluable to the developing field of meteor science at Jodrell Bank. Since this new field was advancing very rapidly, Lovell, Prentice and Porter jointly organised a course on meteor science and radar in Manchester in March 1947.

With Porter's encouragement (2), Lovell was elected to the BAA on 31 December 1947, together with three other members of his new team: J. A. Clegg, J. G. Davies and V. A.

Hughes. Porter proposed their nominations and Prentice seconded them. Mary Almond joined Lovell's team in 1949 and was likewise encouraged to join the BAA.

Lovell and other members of his group spoke at many BAA meetings over the years and published papers in the *Journal*. In this way, the professional research group developing at Jodrell Bank became increasingly connected with the wider amateur astronomical community. These pioneering investigations laid the foundations not only for radar meteor astronomy but also for the wider development of radio astronomy at Jodrell Bank.

Bernard Lovell was appointed Professor of Radio Astronomy at the University of Manchester in 1951—the first such appointment anywhere in the world. To mark the occasion, a photograph of his research group was taken with Jodrell Bank's Searchlight Aerial as a backdrop (Figure 1). The photograph includes five BAA members, among them Lovell himself (V. A. Hughes had moved on from Jodrell by that time).

The lives and work of Lovell, Prentice and Porter have been covered in other publications, for example in books by Lovell (3) (4) and in an earlier paper by the present author: *The Birth of Radar Meteor Astronomy at Jodrell Bank: The Collaboration between Bernard Lovell and Manning Prentice* (1).

This paper examines the lives of the other BAA members who were associated with the meteor programme at Jodrell Bank during the years of the closest links between the BAA and Jodrell, from 1946 to 1951, and who are less well known. It is remarkable that they all went on to make important contributions to science. Moreover, each left behind a scientific legacy extending from the earliest days of radar meteor astronomy at Jodrell Bank. Their achievements illustrate the remarkable versatility of the generation of scientists who transformed British science in the decades following the Second World War.

**Michael William Ovenden (1926–1987)**

Michael William Ovenden (Figure 2) was born in Muswell Hill, London, on 21 May 1926 (5) (6) (7). He developed an early interest in astronomy and was elected to the BAA on 31 March 1943 at the age of sixteen. Shortly afterwards, he attended the June 1943 BAA meeting in London, where he asked a question following a talk on Kenneth Edgeworth's paper *The Evolution of our Planetary System* (8). Over the next few years, he attended many meetings and was frequently heard asking questions and joining in discussions on the topics under consideration.

The following year, at the July 1944 BAA meeting, he presented his first paper, *A Visual Photometer for the Comparison of Stars of Different Colours* (9) (10). Several other papers and notes followed on diverse subjects, including the origin of lunar craters (11), stellar magnetism & spectroscopy (12) (13) (around this time he borrowed BAA instrument 123, a Zöllner spectroscope (14)), solar prominences, quartz clocks (15) and, of course, meteors. He also exhibited drawings of Mars and of eruptive solar prominences at the 1946 and 1947 Exhibition Meetings, respectively.

Given his increasing involvement with the BAA and his evident scientific ability, it was perhaps no surprise that Ovenden became Papers Secretary in 1946. Remarkably, he was only twenty years old and thus the youngest Papers Secretary in the Association's history. He took over the role from Frank Maurice Holborn, who became BAA President the same year. He served as Papers Secretary until 1951. In those days, the practice at BAA meetings was for recently submitted papers to be read, either by the author or by a senior member of the Association. The paper itself was published in the same edition of the *Journal*, or occasionally in a subsequent issue. Other talks were generally notified by prospective speakers on the day of the meeting, sometimes only shortly before it began, allowing topical subjects to be presented and discussed. There was usually no shortage of volunteers to speak, but occasionally Ovenden found himself having to give an impromptu talk, so he generally kept something up his sleeve. Given his wide-ranging interests in astronomy, he never found it difficult to choose a suitable topic.

He was elected a Fellow of the Royal Astronomical Society in 1945. From 1957 to 1966 he served as a Secretary and Council Member of the Society. He was also editor of *The Observatory* from 1951 to 1952.

Ovenden was already an active meteor observer by the mid-1940s and his work was well known to the Meteor Section Director, Manning Prentice. When Prentice was contacted by Nikolai Herlofson of the University of Manchester about supplying observers to visit Jodrell Bank for the 1946 Perseids, Ovenden was at the top of his list (16) (Figure 3). He was already on summer vacation from his studies at Queen Mary College, London, and therefore had time to travel from his London home to Manchester. Later in life, he commented that several of the nights at Jodrell were partially cloudy. He did not mind, however, as he was able to join Lovell at the radar display and watch the echoes from Perseid trains despite the cloud (17).

After completing his first degree at Queen Mary in 1947, he moved to Cambridge, where he was Junior Assistant Observer at the Cambridge Observatories from 1948 to 1952, gaining first an MA and then a PhD. There he immersed himself in practical astronomy, making extensive use of a pulse-counting photoelectric photometer attached to the 15-inch (38 cm) Huggins refractor to study eclipsing binaries such as ζ Aur (18) and GO Cyg (19). He regularly used a spectrohelioscope at the Cambridge Observatories. (20)

The eclectic environment of Cambridge refined his interests in celestial mechanics and stellar dynamics. Cambridge researchers were also beginning work in radio astronomy and they sometimes asked Ovenden for help to identify the optical counterparts of the radio sources they had detected. He found this both challenging and amusing because the positions they supplied were often uncertain by several degrees, due to the low resolution of the radio maps of the time.

In 1953 Ovenden joined the University of Glasgow as a lecturer in astronomy. During the next thirteen years he established himself as an inspiring teacher and active researcher, becoming Senior Lecturer in 1964. His contributions to Scottish astronomy were recognised by his election as a Fellow of the Royal Society of Edinburgh in the same year. He was also closely associated with the Astronomical Society of Glasgow, eventually serving as its President and later Honorary President, thereby playing an important role in fostering amateur astronomy.

His move to Glasgow meant that he attended BAA meetings in London rarely, and he stepped down as Vice-President in 1953. He was invited to present the BAA Christmas Lecture on 4 January 1961 on *The Origin of the Constellations*. He retained his links with the BAA even when he later moved to Canada, writing in 1972: "I follow in the *Journal*, regularly and with interest, the proceedings of an Association which I joined, as an enthusiastic schoolboy, nearly thirty years ago." (21)

Ovenden emigrated to Canada in 1966 to become Professor of Astronomy at the University of British Columbia (UBC), where he was charged with developing a modern astronomy programme. Under his leadership the department expanded significantly, and he was instrumental in establishing observational facilities while strengthening both undergraduate and postgraduate teaching. He also maintained close links with the H. R. MacMillan Planetarium in Vancouver (Figure 4), becoming one of western Canada's best-known communicators of astronomy.

Ovenden's research interests ranged widely across astronomy. He published on stellar photometry and spectroscopy, satellite dynamics, astronomical history, planetary systems and cosmology, but became particularly well known for his investigations into the structure and evolution of the Solar System. He argued that the present planetary system could best be explained if an additional major planet had once existed between Mars and Jupiter before being destroyed, leaving the asteroid belt as its remnants (22) (23). In developing this hypothesis, he sought a physical explanation for the empirical Titius–Bode relation governing planetary distances. Although the idea attracted considerable attention and stimulated extensive discussion, later work showed that the asteroid belt contains far too little mass to represent the remains of a destroyed planet, and the hypothesis is no longer generally accepted. Nevertheless, his work renewed interest in the dynamics and long-term evolution of planetary systems.

Throughout his career, Ovenden exhibited an exceptional ability to explain astronomy to non-specialists. He wrote a number of successful popular books, including *Looking at the Stars* (1957), *Artificial Satellites* (1960) and *Life in the Universe* (1962), together with numerous articles and public lectures. He believed strongly that astronomy should be accessible to everyone and devoted considerable time to education and public engagement.

Ovenden was prepared to challenge accepted ideas and pursue unconventional lines of enquiry whenever he believed the evidence justified them. This intellectual independence earned him both admiration and criticism, but also ensured that his work remained influential and widely discussed.

Ill health forced Ovenden to retire from UBC in 1985. He died suddenly from a heart attack in Vancouver on 15 March 1987, aged sixty. His death brought to an end a career characterised by originality, enthusiasm and a deep commitment to both professional and amateur astronomy. He was remembered not only for his scientific contributions but also for his generosity as a teacher, lecturer and mentor, and for the encouragement he gave to generations of astronomers on both sides of the Atlantic.

It was fitting that, following his death, a memorial service was held at the Vancouver Planetarium for his family and friends.

**Gerald Stanley Hawkins (1928–2003)**

Gerald Stanley Hawkins (Figure 5) was born on 20 April 1928 in Great Yarmouth, Norfolk, where the family home on South Beach Parade directly overlooked the town's Pleasure Beach (24) (25) (26). When he discovered astronomy while at primary school, he borrowed books on the subject from the public library.

It was meteors that captured his imagination: "Many astronomers will admit that their interests began one night many years ago when they sat out under the stars to plot the track of meteors. A few of them, and I am among the number, have retained an interest and have tried to contribute to the growth of meteor astronomy". (27)

He made his first meteor observations from Great Yarmouth and soon joined Prentice's group of observers contributing to the BAA Meteor Section. He later recalled that, "I have watched the Perseids on many occasions—from my home on the east coast of England when I was a member of a team of observers scattered over East Anglia, from a small sailboat bobbing off the coast of the Isle of Wight..... I am always impressed by the event and look forward to the shower as an old friend". (27)

At the start of the Second World War, because of the exposed position of the family home overlooking the sea, Hawkins was evacuated inland away from the threat of German shelling from the North Sea. He settled in the Nottingham area for the duration of the war. There he joined the local astronomical society and systematically observed meteors.

In the summer of 1946, at the age of eighteen, Hawkins' name appeared on Prentice's list of observers for the Perseids in August (Figure 3) (16) (28). It was listed below Ovenden's, who he was scheduled to relieve at Jodrell Bank. The timing was ideal, as he had just left school and was enjoying the long summer vacation before beginning university.

Although he had contributed observations to the Meteor Section for several years (29), Hawkins did not become a member of the BAA until 29 January 1949. Thereafter his name appeared regularly in the *Journal*, particularly in Meteor Section reports. At the June 1949 BAA meeting he presented his first paper on *Meteor Light Curves* (30) (31), which included a mathematical analysis of his own visual observations of meteor trails. The talk was well received and prompted questions from several distinguished members, including W. H. Steavenson, Cicely M. Botley and Gerald Merton. Michael Ovenden, attending in his role as BAA Secretary, also took part in the discussion. A written version of the paper appeared in the *Journal*, in which Hawkins acknowledged

Prentice's assistance in its preparation (31). He also served on the BAA Council during the 1950–51 Session (32).

Hawkins studied physics at the University of Nottingham, graduating in 1949, while also taking a subsidiary degree in mathematics from the University of London. He then returned to Jodrell Bank to begin doctoral studies under Lovell at the University of Manchester, researching radar meteor astronomy.

Hawkins' doctoral research mainly focused on the annual meteor showers. He also observed Bečvář's meteor stream in December 1949, the results of which he presented at the May 1950 BAA meeting (33) together with his co-author, Mary Almond (34). Later in his studies, he worked on the daylight meteor streams of 1949 and 1950 (35) which Lovell and Clegg discovered in 1947 (discussed in the section on Clegg, below).

The rapidly extending radar work at Jodrell combined with a small team meant that members often became engaged in a variety of projects. For example, Hawkins became involved in the project to apply radar in the observation of aurorae. He presented his findings at the February 1952 BAA meeting (36), where Ovenden opened the discussion. The published paper, *Radio Echo Reflections from the Aurora Borealis* (37), described how radar was used during the intense auroral display of the night of 15–16 October 1949 (Figure 6). He demonstrated how the radar could distinguish between different components of the aurora, such as arcs and rays. This was an active period for aurorae, and Hawkins also made visual observations of the display of 30 May 1949 from Radcliffe on Trent and the “Great Aurora” on 17 April 1947 (38).

During his doctoral research he continued visual meteor observing, describing how he watched the persistent train of a bright (magnitude –1.5) Perseid become distorted over several seconds on 14 August 1950 (39).

Hawkins received his PhD from the University of Manchester in 1952 (the University later awarded him a DSc in 1963). He then worked briefly on classified defence research for Ferranti before emigrating to the United States in 1954. There he joined the Harvard-Smithsonian Observatory in Cambridge, Massachusetts, as a research astronomer. Drawing upon his experience at Jodrell Bank, he also became a Research Associate with Fred Whipple's Harvard Radio Meteor Project at Harvard University, as well as a Senior Associate at the United States Air Force Cambridge Research Laboratory.

In 1957 he was appointed Professor of Astronomy at Boston University, where he later became Chairman of the Department of Astronomy. Between 1969 and 1971 he served as Dean of the Faculty of Liberal Arts at Dickinson College before returning to research and science administration, including work for the United States Information Agency. Throughout his career he remained closely associated with the Smithsonian Institution and was an energetic advocate of public understanding of science. He became a United States citizen in 1965.

Hawkins' enduring scientific legacy arose from a question that had fascinated visitors to Stonehenge for centuries: was the monument deliberately aligned with the heavens? During a visit to Stonehenge in the early 1950s he became intrigued by the possibility that its layout possessed a more sophisticated astronomical purpose than had previously been recognised. Returning to the problem a decade later, he employed one of the world's earliest large electronic computers—an IBM 7090—to analyse the positions of the stones and earthworks. In 1963 he published his results in *Nature*, (40) arguing that many of the monument's alignments were intentional and that Stonehenge had functioned as a prehistoric astronomical observatory capable of predicting eclipses.

These ideas were developed in his book, *Stonehenge Decoded* (1965) (41). It became an international bestseller and permanently altered public perceptions of Stonehenge. Hawkins argued that the monument was not simply a ceremonial structure but an ingenious astronomical instrument—a "Neolithic computer"—whose builders possessed a far more sophisticated understanding of celestial motions than had previously been imagined. Although many archaeologists criticised his conclusions, particularly his claims regarding eclipse prediction, the book inspired widespread interest in the discipline now known as archaeoastronomy. Later generations of researchers rejected some of Hawkins' more ambitious interpretations but confirmed that astronomical alignments formed an important element in the planning of many prehistoric monuments.

Following the success of *Stonehenge Decoded*, Hawkins broadened his investigations to other ancient sites. He studied the stone circles at Callanish on Lewis, the Temple of Amun at Karnak in Egypt, the Nazca Lines of Peru and prehistoric cave art, applying quantitative astronomical methods to each. Although these investigations met with varying degrees of acceptance, they reflected his lifelong conviction that astronomy had played a central role in many ancient cultures. In later years he also examined the geometry of crop circles, concluding that they were the work of human designers rather than natural phenomena or extraterrestrial visitors.

Alongside his research, Hawkins was a gifted communicator. He wrote eleven books and more than 150 scientific papers covering astronomy, meteors, cosmology, astrobiology and the history of astronomy. His other books included *Splendor in the Sky* (1961), *Beyond Stonehenge* (1973), *Mindsteps to the Cosmos* (1983) and, completed shortly before his death, *Stonehenge, Earth and Sky* (2004). He also appeared in television documentaries and was widely recognised for his ability to explain complex scientific ideas to a broad audience. His contributions to science communication were recognised by awards from Boston University, the Smithsonian Institution and the U.S. National Academy of Sciences.

Those who knew Hawkins remembered him as imaginative, energetic and intellectually fearless. He relished challenging accepted ideas and was prepared to pursue unconventional hypotheses wherever he believed the evidence led. His willingness to apply new computational techniques to ancient archaeological problems was ahead of its time and, although many of his conclusions remain debated, his methodological innovations transformed both archaeology and the history of astronomy.

Hawkins died suddenly of a heart attack at his home in Virginia, on 26 May 2003, aged seventy-five.

**John Atherton Clegg (1913–1987)**

John Atherton Clegg (Figure 7) was born on 15 May 1913 (42). He studied physics at the University of Manchester, graduating in 1935. He began his career as a schoolteacher but with the onset of war, became a Scientific Officer at the Telecommunications Research Establishment (TRE) in 1940. There he worked on radar aerial design. It was at TRE that he first met Bernard Lovell, who was also engaged in radar research there.

By 1946, Lovell's radar programme was showing great promise, and Patrick Blackett, head of the Manchester Physics Department, encouraged him to recruit assistance (3). Admiring Clegg's work at TRE and knowing that he wished to pursue a research career, Lovell soon recruited him to Jodrell Bank. Clegg brought with him extensive expertise in radar aerial design, which was key to designing and constructing the famous Searchlight Aerial (3) (4).

The Searchlight Aerial (Figure 1, 8, 9) proved crucial to the next phase of the Jodrell meteor programme. Lovell wished to direct his radar aerials towards any part of the sky during a meteor shower, both towards the radiant and at right angles to it. To achieve this, he acquired a surplus Army searchlight mounting, dispensing with the reflector but retaining its robust altazimuth mount. This proved ideally suited to supporting an array of five seven-element Yagi aerials constructed by Clegg.

The Searchlight Aerial was commissioned in time for the Giacobinid meteor shower of 9 October 1946. Also known as the Draconids, Prentice anticipated that the event might develop into a meteor storm. Sure enough, meteor rates briefly reached around 10,000 per hour, both visually and by radar (43). The decisive experiment took place close to the time of maximum activity. When the Searchlight Aerial was directed towards the radiant, the number of radar echoes fell dramatically. Rotating the aerial through 90° restored the echo rate, demonstrating that the radar was detecting specular reflections from ionised meteor trains rather than from the meteoroids themselves.

Lovell later wrote: "The Searchlight Aerial survived for many years and became the aerial system for a variety of subsequent research programmes." (3) Eventually the aerial was superseded, but the rusty remains of the searchlight base survive as a reminder of an epic night in the early history of Jodrell Bank. In 1947 Lovell and Clegg presented their findings to the Royal Astronomical Society (3).

In addition to the ex-Army searchlight, Lovell accumulated substantial quantities of other surplus wartime equipment during 1946, often at little or no cost (3). The challenge was collecting it from around the country and transporting it to Jodrell. Clegg proved just the man for the job. One major acquisition was a large ex-Army *Park Royal* truck (Figure 9) packed with radar equipment. However, while attempting to drive it across the Jodrell Bank observing field, Clegg became hopelessly bogged down, the vehicle sinking to its axles in the Cheshire clay. There it remained for several years, serving as Lovell's first headquarters.

One morning, during observations of the η Aquariids in May 1947, Clegg and Lovell left the radar operating after dawn had broken. To their surprise, meteor activity continued well into daylight hours. They had inadvertently discovered the phenomenon of daytime meteor showers, opening another productive avenue of research (44) (45) (46). Further daytime streams were subsequently identified during the summer and in the years that followed (46).

Clegg's next project was to design and build Jodrell's 218-foot (66 m) Transit Telescope, consisting of a wire-mesh parabolic reflector suspended from a circular framework of scaffolding (Figure 10, 11). When completed in late 1947, the telescope was the world's largest radio telescope. Rather pragmatically, its diameter was determined by the space available between the *Park Royal* truck and the hedge at the edge of the field, while its curvature was fixed by the height of the ladders that could reach the outer edge of the reflector. This also meant that the central pole supporting the receiver-transmitter had to be 126 feet (38 m) high. Its construction was a genuine team effort involving Lovell, Clegg, and volunteers including Lovell's wife and children (3).

The instrument led to a series of important discoveries, including the first definitive detection of radio emission from the Andromeda Galaxy (M31) in 1950 by Robert Hanbury Brown and Cyril Hazard (47) (Figure 10) thereby establishing it as the first confirmed extragalactic radio source. It also detected radio emission from the remnant of Tycho's Supernova of 1572 before the source had been identified optically (48).

During 1950 Clegg was involved in the calculations of the beam shape and power gain for a new radio telescope that Lovell was planning: the *Mark I* Radio Telescope (Figure 12).

Clegg remained at Jodrell Bank until 1951. The following year, Lovell and Clegg published their classic book *Radio Astronomy* (49).

Clegg's scientific interests increasingly shifted towards geomagnetism and the magnetic properties of rocks. Joining the research group led by Blackett, he became one of the principal figures in the emerging science of palaeomagnetism. He helped develop increasingly precise techniques for measuring the remanent magnetisation preserved in ancient rocks.

Clegg and his colleagues showed that British Triassic rocks provided some of the earliest quantitative evidence that the Earth's continents had changed both latitude and orientation over geological time (50). At a period when Alfred Wegener's theory of continental drift remained controversial, these measurements provided powerful support for the idea that continents had migrated across the Earth's surface. The work became an important step towards the eventual acceptance of plate tectonics during the 1960s.

Clegg also became one of the pioneers of archaeomagnetism, demonstrating that archaeological materials such as kilns, hearths and fired pottery acquire a permanent magnetic signature as they cool. Because the direction of the Earth's magnetic field changes slowly through time, these magnetic records can be used to determine the age of archaeological structures. His work helped establish archaeomagnetism as a practical scientific dating technique.

Clegg died on 28 August 1987. His scientific legacy extends across several disciplines including astronomy, Earth science and archaeology.

**John Grant Davies (1924–1988)**

John Grant Davies (Figure 1, 13) was born in Croydon on 5 November 1924 (51). He studied Mechanical Sciences at Cambridge, graduating with first-class honours in 1944. He then worked at the Royal Aircraft Establishment before joining Lovell's team at Jodrell Bank in January 1947 to undertake research for a PhD.

Lovell assigned Davies to investigate meteor velocities. It was generally accepted among astronomers that shower meteors were associated with cometary debris and that the resulting meteoroids therefore followed elliptical orbits around the Sun. During the 1940s, however, a more contentious debate developed concerning sporadic meteors. Some researchers argued that a significant proportion of these objects followed hyperbolic trajectories and therefore originated from outside the Solar System.

The hyperbolic interpretation was strongly advocated by E. J. Öpik, but challenged by J. G. Porter and Manning Prentice (3). It was realised that precise measurements of sporadic meteor velocities could distinguish between normal Solar System meteors and genuinely interstellar objects: meteors in hyperbolic orbits would require heliocentric velocities exceeding 42.2 km/s.

Davies' careful experimental work, together with the design of specialised equipment, demonstrated that the velocities of sporadic meteors were below 42.2 km/s and therefore consistent with elliptical, rather than hyperbolic, orbits.

As Jodrell Bank expanded during the 1950s, Davies became increasingly involved in the development and scientific exploitation of its growing range of radio telescopes. He played a major role in designing the motor control system for the *Mark 1* radio telescope that allowed it to track celestial objects across the sky. With the telescope (Figure 14) finally commissioned in 1957, his interests turned to the observation of radio emission from the Galaxy.

Davies was among the first generation of astronomers to exploit the unique capabilities of the *Mark 1* telescope in investigating the distribution of neutral hydrogen, ionised gas and radio continuum emission throughout the Milky Way. His observations made significant contributions to understanding the large-scale structure of the Galaxy and the physical properties of the interstellar medium.

On 12 September 1959, at the request of the Soviet Union, the Lunik II spacecraft was tracked to the Moon using the *Mark 1* telescope. Davies measured the Doppler shift produced by the Moon's gravitational field—a measurement that convinced sceptical American scientists that the Soviet mission had successfully reached the lunar surface and which the U.S. space programme later described as a "stroke of genius" (51).

Davies also played a major role in the development of the University of Manchester's radio astronomy programme. Appointed Professor of Radio Astronomy in 1966, he supervised numerous postgraduate students and became widely respected as an inspiring teacher and careful scientific mentor.

Davies died on 16 September 1988 at the age of sixty-three. His career bridged the pioneering era of post-war radar astronomy and the maturity of modern radio astrophysics.

**Victor A. Hughes (1925–2001)**

Victor Alfred Hughes (Figure 15) was born on 20 April 1925 (52). He graduated with a degree in physics from the University of Manchester in 1944 and, for the remainder of the Second World War, was seconded to the Telecommunications Research Establishment at Malvern, where he acquired extensive experience of radar systems. This wartime work proved decisive in shaping his scientific career, introducing him to the techniques that underpinned the development of radio astronomy in the immediate post-war years.

Hughes returned to Manchester in September 1947 as one of Lovell's postgraduate students at Jodrell Bank. His research initially concentrated on radar observations of the recently discovered daylight meteor showers (44). But in November 1947 he began his main work: radio observations with the new 218-foot (66 m) Transit Telescope. This was Lovell's first opportunity for proper work with the new telescope. Hughes produced a broad-beam radio map of the sky that revealed a prominent region of emission in Cygnus.

Hughes presented his thesis in the autumn of 1949. Following appointments as Principal Scientific Officer at the Royal Radar Establishment, Malvern, and later at the Radio and Space Research Station, Slough, Hughes emigrated to Canada in 1963 to join Queen's University, Kingston, as Professor of Physics. There he played a leading role in establishing radio astronomy as a major research activity.

During the latter part of his career Hughes concentrated on the radio astronomy of the Galaxy. Using major international facilities, including the Algonquin Radio Observatory, the Westerbork Synthesis Radio Telescope, MERLIN and the Very Large Array, he investigated Galactic structure, interstellar molecules and regions of active star formation. He became particularly interested in the complex radio sources associated with young stellar objects, carrying out extensive studies of the Cepheus A star-forming region. His observations revealed rapid radio variability and led him to propose that powerful magnetic reconnection events accompanied the earliest stages of stellar evolution. Although some aspects of this interpretation remained controversial, his work stimulated considerable discussion concerning the role of magnetic fields in star formation. One of the compact radio sources in Cepheus A, discovered jointly with collaborators, became known as the Hughes–Wouterloot object (53).

Hughes was equally committed to the development of astronomy in Canada. He was an early and enthusiastic supporter of the Canadian Astronomical Society (CASCA) in which Michael Ovenden was also very active. His scientific achievements were further recognised by the award of the degree of Doctor of Science (DSc) by the University of Manchester.

Hughes died on 24 April 2001, shortly after his seventy-sixth birthday.

**Mary Almond (1928–2015)**

Mary Almond (Figure 1, 16) was born in Manchester on 2 January 1928. She entered the University of Manchester in 1946 to study physics, where she was taught by Bernard Lovell.

At the end of their first year of physics lectures, Lovell asked whether any of the male students would be interested in spending some time at Jodrell Bank over the summer, helping with tasks such as digging trenches, mixing concrete, and other physical work. Almond approached him afterwards and asked, (54) “Would there be anything for girls to do at Jodrell?” Lovell replied that he was sure there would be.

Almond, along with a female friend who was in the same year, spent two weeks working at Jodrell Bank, living at Alderley Edge in a caravan belonging to Almond’s former physics teacher and cycling to Jodrell Bank each day. There, they sandpapered rust from the mount of the Searchlight Aerial. During these two weeks, Almond also witnessed Prentice carrying out visual meteor observations.

During the following summer, 1948, Almond returned to Jodrell Bank, but this time she camped in a tent alongside other physics students, all of whom were male (54).

Almond graduated with a 2:1 physics degree in 1949 and stayed on to undertake doctoral research at Jodrell Bank under John G. Davies, completing her PhD in 1952. She was thus a contemporary of Hawkins in that their research spanned the same years.

Almond investigated the velocity distribution of sporadic meteors and the possibility of interstellar meteors (34) (50) (55) (56) (57) (58). She also collaborated with Hawkins on radar observations of the principal night-time meteor streams, work that helped establish radar techniques as a powerful tool for investigating meteoroid orbits. As mentioned earlier, they both presented their results on radar observations of Bečvář’s meteor stream at the May 1950 BAA meeting (34).

Since this was a relatively new field, finding an external examiner for Almond's thesis proved difficult until Lovell persuaded Erwin Finlay-Freundlich (1885 –1964) from St Andrews University to undertake the task. She was awarded her PhD in 1952.

Although she received an offer to join Ferranti after completing her doctorate, Almond instead joined Blackett's palaeomagnetism research group. Working with John Clegg, she contributed to pioneering studies of the remanent magnetism preserved in Triassic rocks which was discussed in the section of Clegg, above. (59) (50)

Almond married around this time and moved with Blackett to Imperial College London when he was appointed professor there. She remained at Imperial until 1954, when she left on a career break to have children.

Returning to work, she first taught science in a secondary school before retraining in the emerging discipline of computer programming. She subsequently became one of Britain's earliest lecturers in computer science, teaching mathematics and computing at the University of London and later at the University of Manchester before joining the Open University. There, she spent many years developing and teaching courses in mathematics and computing, remaining academically active until her retirement in 2008 at the age of eighty.

Mary Almond died on 26 April 2015, aged eighty-seven. Few scientists have contributed significantly to so many emerging disciplines: from the birth of radio astronomy and the study of meteors, through the development of palaeomagnetism as evidence for continental drift, to the establishment of computer science as a university subject.

**Conclusion**

The development of radar meteor astronomy at Jodrell Bank in the years following the Second World War was the result of a unique convergence of scientific opportunity, technical expertise and collaboration between professional and amateur astronomers. Bernard Lovell's chance discovery of radar echoes from meteor trains in 1945 opened a new field of research, but the rapid progress that followed depended on a much wider community. The observational experience of Manning Prentice and his BAA Meteor Section team, together with the scientific and technical abilities of Lovell's research group, provided an essential foundation for interpreting and extending these early discoveries.

The individuals associated with this pioneering programme demonstrate the extraordinary breadth of talent that emerged from Jodrell Bank during this formative period. Michael Ovenden and Gerald Hawkins started out as enthusiastic young meteor observers within the BAA but went on to distinguished scientific careers. One wonders to what extent the remarkable opportunity of working alongside Lovell in 1946 gave each impetus to pursuing science as a career. John Clegg and Victor Hughes applied their wartime radar experience to the practical challenges of building and exploiting the first radio telescopes, contributing not only to the establishment of radio astronomy but also, in Clegg's case, to the revolutionary development of palaeomagnetism and archaeomagnetism. John Davies helped transform radar meteor observations into precise quantitative science before making major contributions to Galactic radio astronomy. Mary Almond's career illustrates equally the versatility of this generation, moving from pioneering research in meteor astronomy and palaeomagnetism to becoming one of Britain's early contributors to computer science.

A striking feature of this group is the extent to which their later achievements reached far beyond their original research fields. Their careers crossed disciplinary boundaries at a time when new scientific techniques were transforming astronomy, Earth science and computing. The same qualities that characterised the early work at Jodrell Bank—experimental ingenuity, collaboration, and a willingness to explore unfamiliar territory—remained evident throughout their subsequent careers. Even where individual ideas were later revised, as in some of the interpretations of planetary formation or ancient astronomical sites, their willingness to apply new methods and ask ambitious questions had a lasting influence.

The close relationship between Jodrell Bank and the BAA between 1946 and 1951 therefore represents more than a historical connection between an emerging research institution and an amateur society. It illustrates how scientific progress can arise from cooperation between different communities, combining practical observation, theoretical insight and engineering skill. The members of this pioneering network helped establish radar meteor astronomy as a new discipline and contributed to some of the most significant scientific developments of the second half of the twentieth century. Their achievements stand as a testament to the creativity and versatility of the generation that helped transform post-war British science.

**Acknowledgements**

I am grateful to the many people who have generously assisted with the research for this paper, in particular James Dawson and Kate Bond, Assistant Archivist at the RAS, for their help in locating information and clarifying historical details. I also thank the librarians and archivists at the John Rylands Library, the University of Manchester, for their invaluable assistance in accessing the Bernard Lovell Archive (references cited as "Letter" are from this archive).

## Works Cited

1. ***Shears J., The Birth of Radar Meteor Astronomy at Jodrell Bank: The Collaboration between Bernard Lovell and Manning Prentice, J. Brit. Astron. Assoc., in press (2026).***

2. ***Letter Lovell to Prentice, 13 November 1947.***

3. ***Lovell A.C.B., Astronomer by Chance, Macmillan London Ltd (1991).***

4. ***Lovell A.C.B., The Story of Jodrell Bank, Oxford University Press, London (1968).***

5. ***Roy A.E., Quart. J. Roy. Astron. Soc., 29, 91-91 (1988).***

6. ***Roy A.E., J. Brit. Astron. Assoc., 98, 53 (1987).***

7. ***Roy A.E., The Observatory, 108, 21-31 (1988).***

8. ***J. Brit. Astron. Assoc., 53(6), 172 (1943).***

9. ***J. Brit. Astron. Assoc., 54(8), 156-157 (1944).***

10. ***Ovenden M.W., J. Brit. Astron. Assoc., 54(8), 172-174 (1944).***

11. ***Ovenden M.W, J. Brit. Astron. Assoc., 56(5), 95 (1946).***

12. ***J. Brit. Astron. Assoc., 54(8), 156-157 (1944).***

13. ***Ovenden M.W., J. Brit. Astron. Assoc., 54(8), 172-174 (1944).***

14. ***J. Brit. Astron. Assoc., 58(7), 2517 (1948).***

15. ***Ovenden M.W., J. Brit. Astron. Assoc., 60(1), 31-36 (1949).***

16. ***Letter Prentice to Herlofson, 27 June 1946.***

17. ***Ovenden M.W., Oral History, https://open.library.ubc.ca/collections/citraudio/items/1.0135218?o=11. (1972). Accessed 20 July 2026.***

18. ***Beer A. and Ovenden M.W., Astrophys. J., 113, 439 (1951). DOI: 10.1086/145413.***

19. ***Ovenden M.W., MNRAS, 114, 569–582 (1954). DOI: 10.1093/mnras/114.5.569.***

20. ***Ovenden M.W., J. Brit. Astron. Assoc., 60(2), 51-55 (1950).***

21. ***Ovenden M.W., J. Brit. Astron. Assoc., 82(2), 106-111 (1972).***

22. ***Ovenden M.W., Nature, 239, 508-509 (1972).***

23. ***Ovenden M.W., Vistas in Astronomy, 18, 473-496 (1974). DOI: 10.1016/0083-6656(75)90128-2.***

24. ***Krupp E. C., "Obituary: Gerald S. Hawkins, 1928-2003". ResearchGate. Retrieved 10 July 2026 (2003).***

25. ***Pitts M., "Gerald Hawkins. Astronomer who claimed Stonehenge was a computer", The Guardian (24 July 2003).***

26. ***"Gerald Stanley Hawkins". Biographical notes from Dickinson College, USA, https://archives.dickinson.edu/encyclopedia/gerald-stanley-hawkins-1928-2003 (2005) Accessed 20 July 2026.***

27. ***Hawkins G.S., Splendor in the Sky, Harper, New York (1961).***

28. ***Letter Prentice to Lovell, 4 August 1945.***

29. ***J. Brit. Astron. Assoc., 58(7), 242 (1948).***

30. ***J. Brit. Astron. Assoc., 59(8), 240-241 (1949).***

31. ***Hawkins G.S, J. Brit. Astron. Assoc., 59(8), 245-248 (1949).***

32. ***J. Brit. Astron. Assoc., 61(1), 19 (1950).***

33. ***J. Brit. Astron. Assoc., 60(7), 184-185 (1950).***

34. ***Hawkins G.S. and Almond M., J. Brit. Astron. Assoc., 60(8), 251 (1950).***

35. ***Aspinall A. & Hawkins G.S., MNRAS, 111, 18 (1981). DOI: 10.1093/mnras/111.1.18.***

36. ***J. Brit. Astron. Assoc., 60(5), 125-127 (1950).***

37. ***Hawkins G.S. and Aspinall A., J. Brit. Astron. Assoc., 60(5), 130-135 (1950).***

38. ***J. Brit. Astron. Assoc., 57(4), 171 (1947).***

39. ***Hawkins G.S., J. Brit. Astron. Assoc., 61(3), 75 (1951).***

40. ***Hawkins G.S., Nature, 200, 306–308 (1963).***

41. ***Hawkins G.S., Stonehenge Decoded, Doubleday (1965).***

42. ***Quart. J. Roy. Astron. Soc., 29, 3, 403 - 407 (1988).***

43. ***Prentice J.P.M., Occasional notes for my friends and other papers, Books 1 & 2, Stowprint (1980).***

44. ***Clegg J. A., Hughes V. A. and Lovell A. C. B., MNRAS, 107, 369 (1947). DOI: 10.1093/mnras/107.4.369.***

45. ***Porter J.G., J. Brit. Astron. Assoc., 58(2), 50-53 (1948).***

46. ***Aspinall A., Clegg J. A. and Lovell A. C. B., MNRAS, 109, 352 (1949). DOI: 10.1093/mnras/109.3.352.***

47. ***Hanbury Brown R. & Hazard C., Nature, 166, 901-902 (1950). DOI: 10.1038/166901a0.***

48. ***Hanbury Brown R. & Hazard C., Nature, 170, 364-365 (1952). DOI: 10.1038/170364a0.***

49. ***Lovell A.C.B. and Clegg J.A., Radio Astronomy, Chapman & Hall, (1952).***

50. ***Clegg J.A., Almond M. and Stubbs P.H.S., Philosophical Magazine, 45, 583-598 (1954). DOI: https://doi.org/10.1080/14786440608520463.***

51. ***Lovell A.C.B., Quart. J. Roy. Astron. Assoc., 30, 365-369 (1989).***

52. ***Henriksen R.M., Bulletin AAS, 33, 4 (2001).***

53. ***Hughes V. A. & Wouterloot J. G. A., Astrophys. J., 276, 204 (1984). DOI: 10.1086/161603.***

54. ***Merchant P., "National Life Stories: an oral history of British science C1379/38, Dr Mary Almond Interviewed by Dr Paul Merchant", British Library. Accessed 13 February 2017.***

55. ***Almond M., Davies J.G. and Lovell A.C.B. The Observatory, 70, 112–113 (1950).***

56. ***Almond M., Davies J.G., Lovell A.C.B. , MNRAS, 11, 585–608 (1950). DOI: https://doi.org/10.1093/mnras/111.6.585.***

57. ***Almond M., Davies J.G., Lovell A.C.B., MNRAS, 112, 21–29 (1950). DOI:10.1093/mnras/112.1.21.***

58. ***Almond M., Davies J.G., Lovell A.C.B, 113, 411–427 (1953). DOI: 10.1093/mnras/113.4.411.***

59. ***Almond M., Clegg J.A., Jaeger J.C., Philosophical Magazine, 1, 771–782 (1956). DOI: https://doi.org/10.1080/14786435608238152.***

60. ***The names cited are from: Spencer R., Galaxies 2017, 5, 68 (2017). https://doi.org/10.3390/galaxies5040068.***

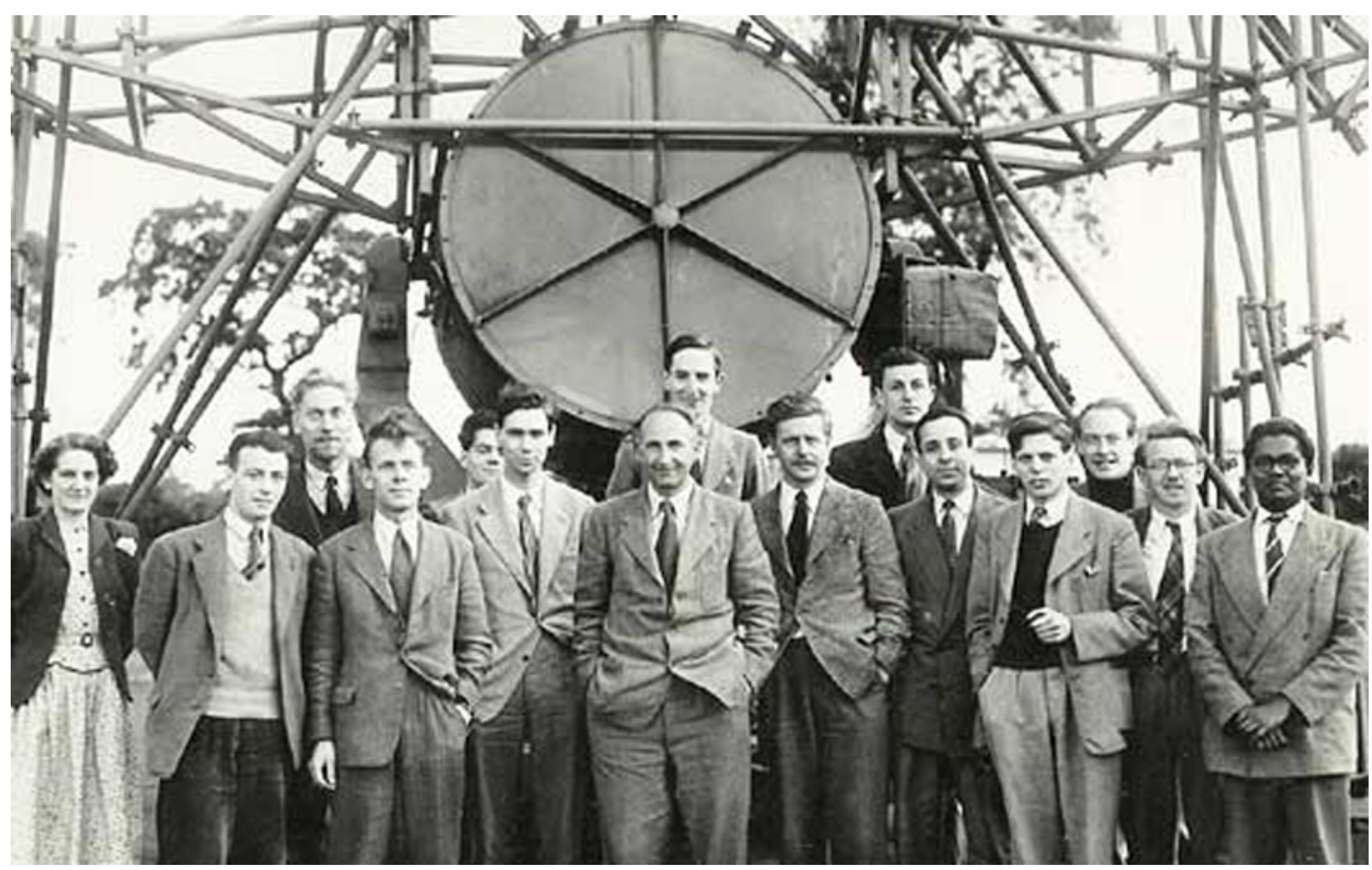

Figure 1: Lovell's team by the Searchlight Aerial at Jodrell Bank, 1951 (the University of Manchester)

People mentioned in this paper: Bernard Lovell (front, centre), J.G. Clegg on his left and J.G. Davies on his right. Gerald Hawkins is behind Davies, partially obscured. Mary Almond is on the extreme left.

L to R: (60) Mary Almond, Cyril Hazard, Roger Jennison, Stanley Greenhow, Gerald Hawkins, John Davies, Bernard Lovell, Gordon Little, John Clegg, Alan Maxwell, Ismail W.B. Hazzaa, Tom Kaiser, Sandy Murray, Tim Close and Mrinal Das Gupta. D. Greening and Robert Hanbury Brown are missing.

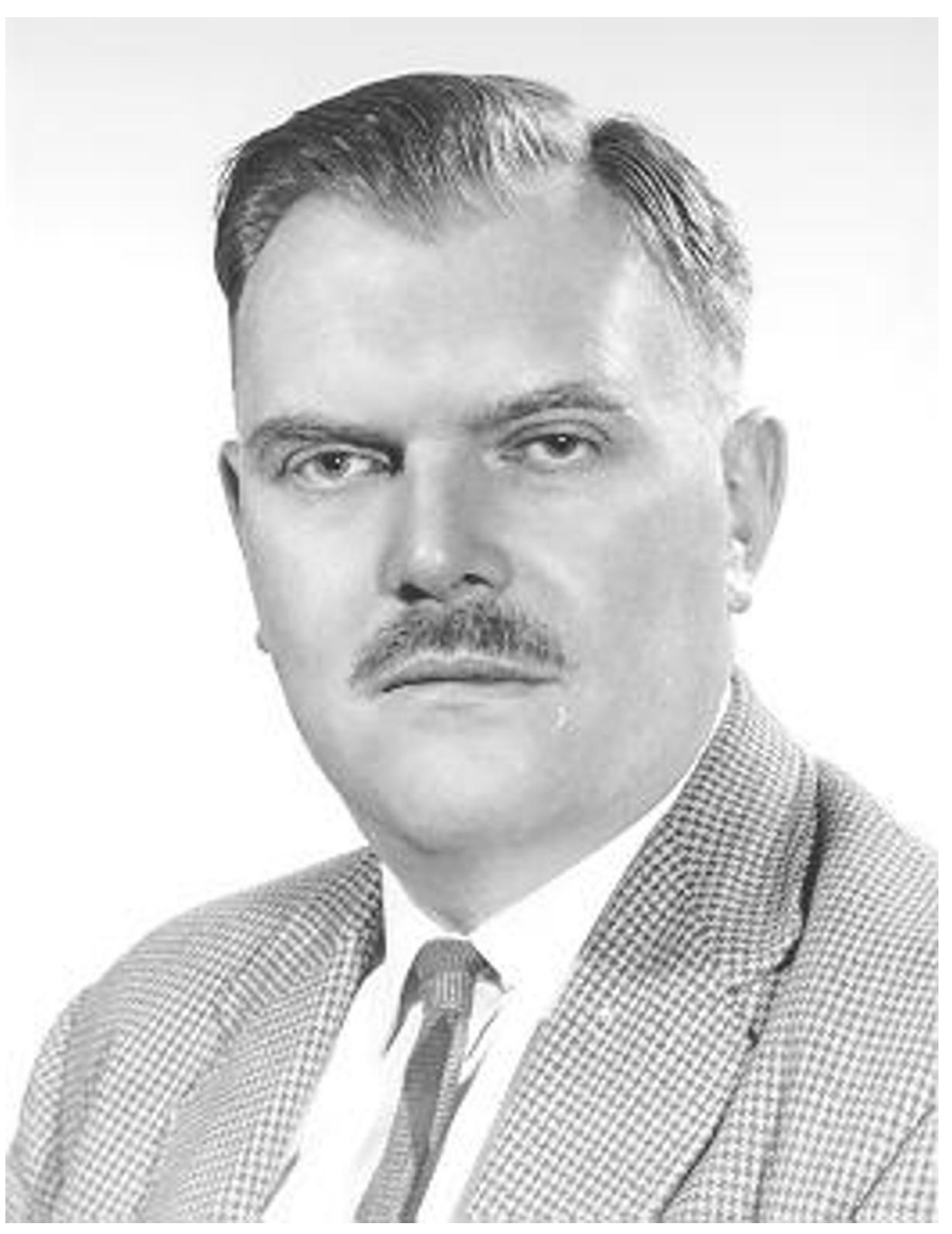

Figure 2: Michael William Ovenden in 1967 (University of British Columbia)

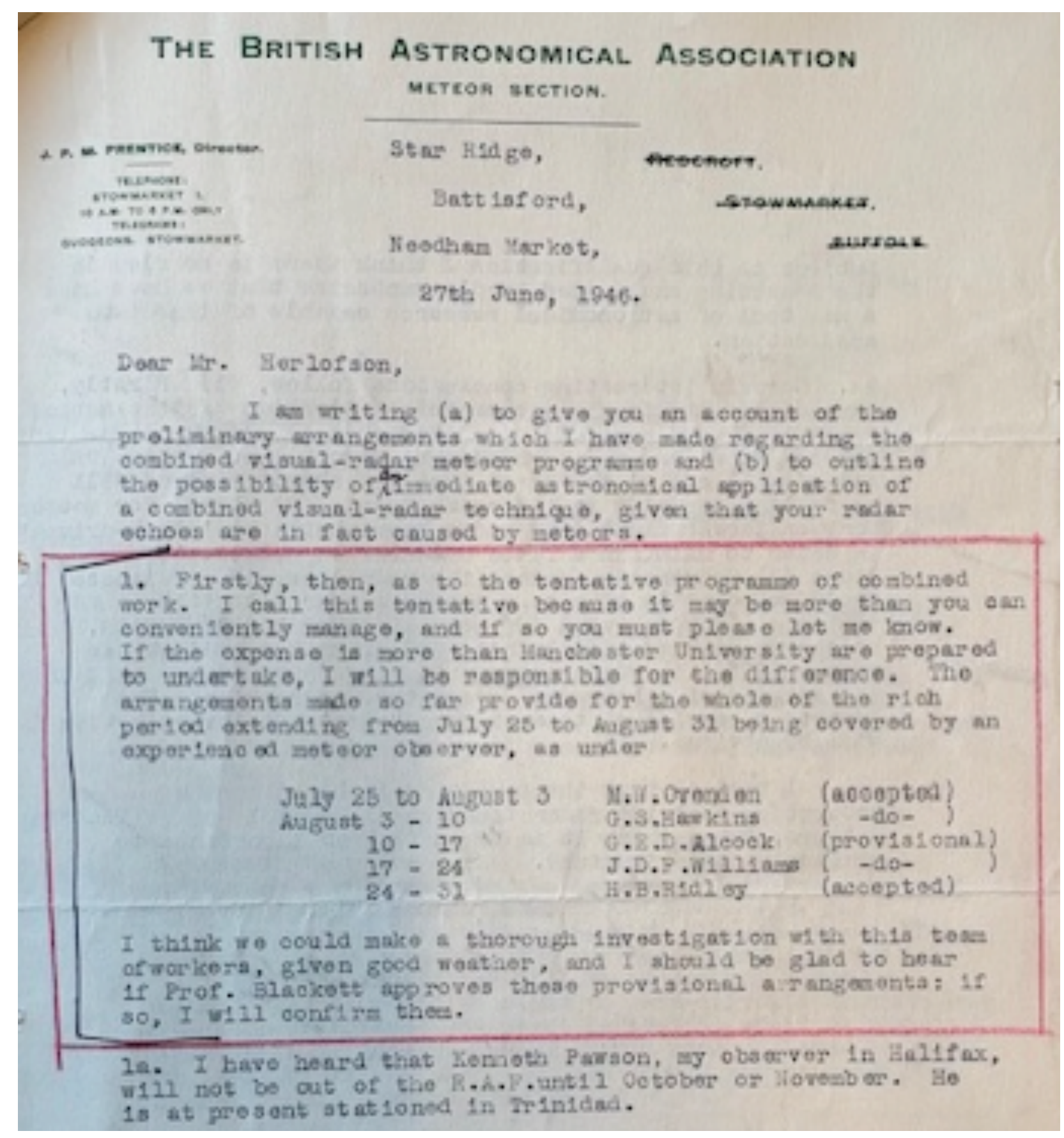

THE BRITISH ASTRONOMICAL ASSOCIATION
METEOR SECTION.

J. P. M. PRENTICE, Director.

Star Ridge,
Battisford,
Needham Market,

~~REDCROFT,~~
~~STOWMARKET,~~
~~SUFFOLK~~

27th June, 1946.

Dear Mr. Herlofson,

I am writing (a) to give you an account of the preliminary arrangements which I have made regarding the combined visual-radar meteor programme and (b) to outline the possibility of an immediate astronomical application of a combined visual-radar technique, given that your radar echoes are in fact caused by meteors.

1. Firstly, then, as to the tentative programme of combined work. I call this tentative because it may be more than you can conveniently manage, and if so you must please let me know. If the expense is more than Manchester University are prepared to undertake, I will be responsible for the difference. The arrangements made so far provide for the whole of the rich period extending from July 25 to August 31 being covered by an experienced meteor observer, as under

| | | |
|---|---|---|
| July 25 to August 3 | M.W.Ovenden | (accepted) |
| August 3 - 10 | G.S.Hawkins | ( -do- ) |
| 10 - 17 | G.E.D.Alcock | (provisional) |
| 17 - 24 | J.D.F.Williams | ( -do- ) |
| 24 - 31 | H.B.Ridley | (accepted) |

I think we could make a thorough investigation with this team ofworkers, given good weather, and I should be glad to hear if Prof. Blackett approves these provisional arrangements: if so, I will confirm them.

1a. I have heard that Kenneth Pawson, my observer in Halifax, will not be out of the R.A.F.until October or November. He is at present stationed in Trinidad.

Figure 3: Letter from Prentice (16) proposing Meteor Section observers to visit Jodrell Bank for the 1946 Perseids, including Ovenden and Hawkins (the University of Manchester)

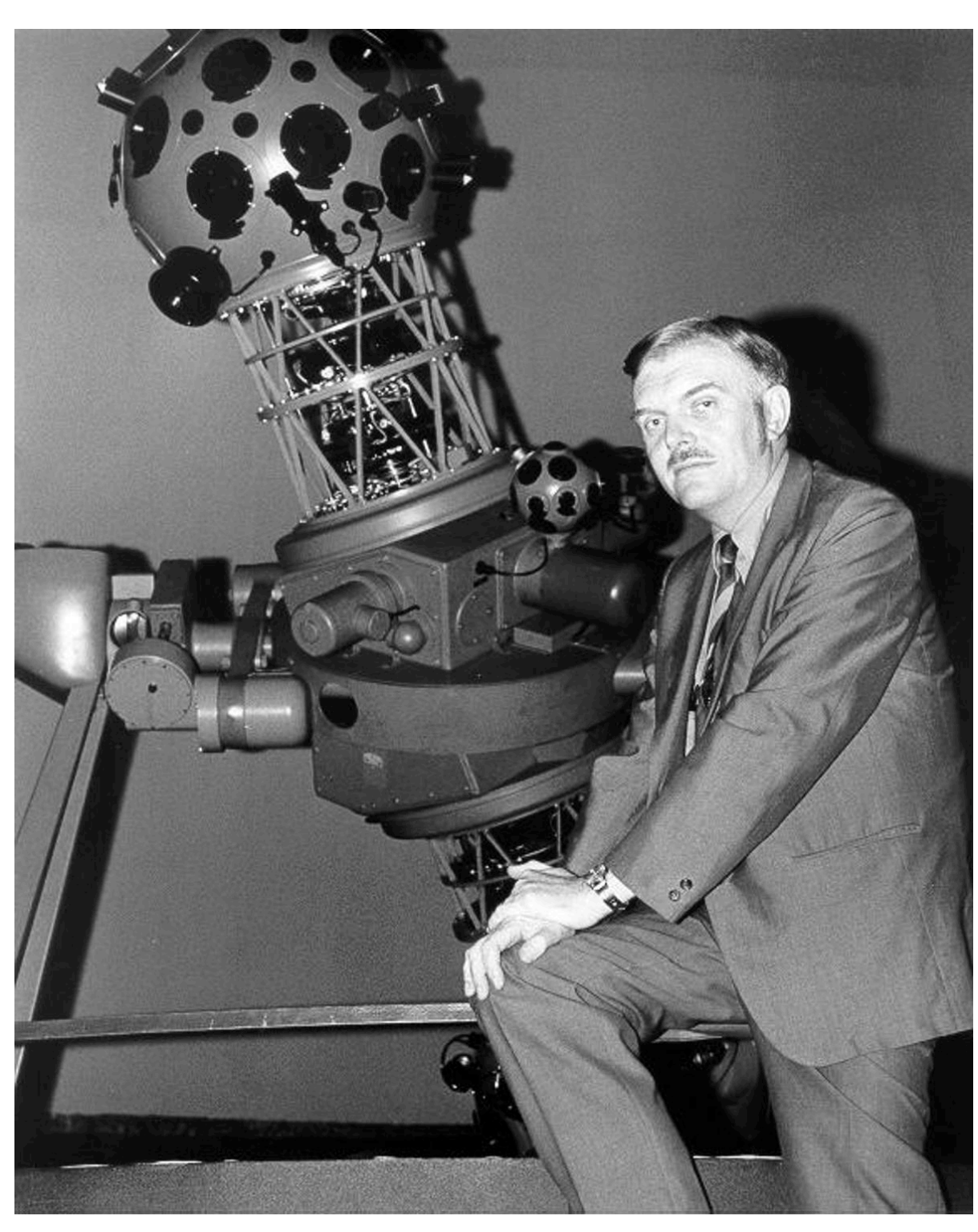

Figure 4: M.W. Ovenden at the H. R. MacMillan Planetarium in Vancouver in the 1970s (University of British Columbia)

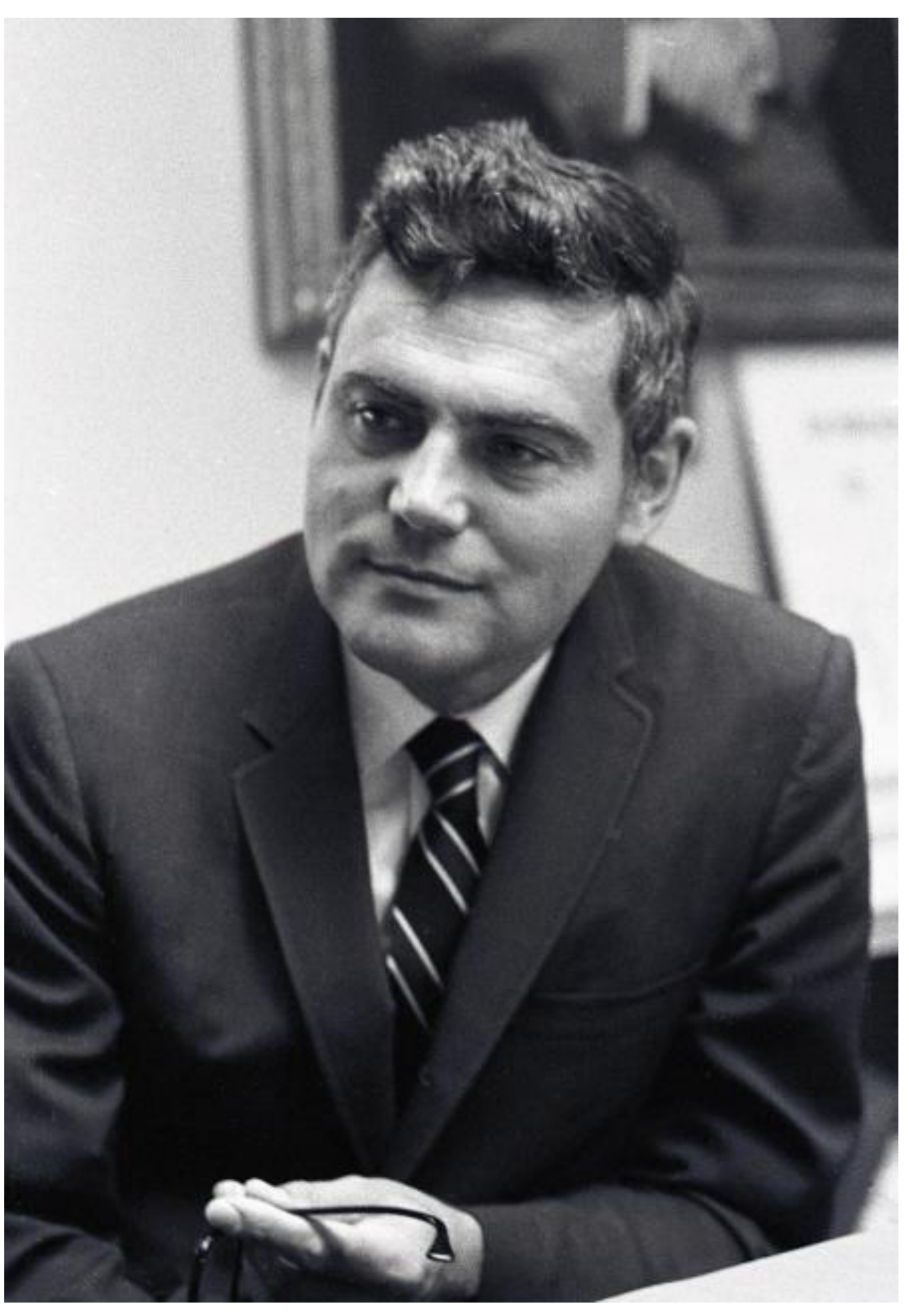

Figure 5: Gerald Stanley Hawkins in 1969 (Dickinson College)

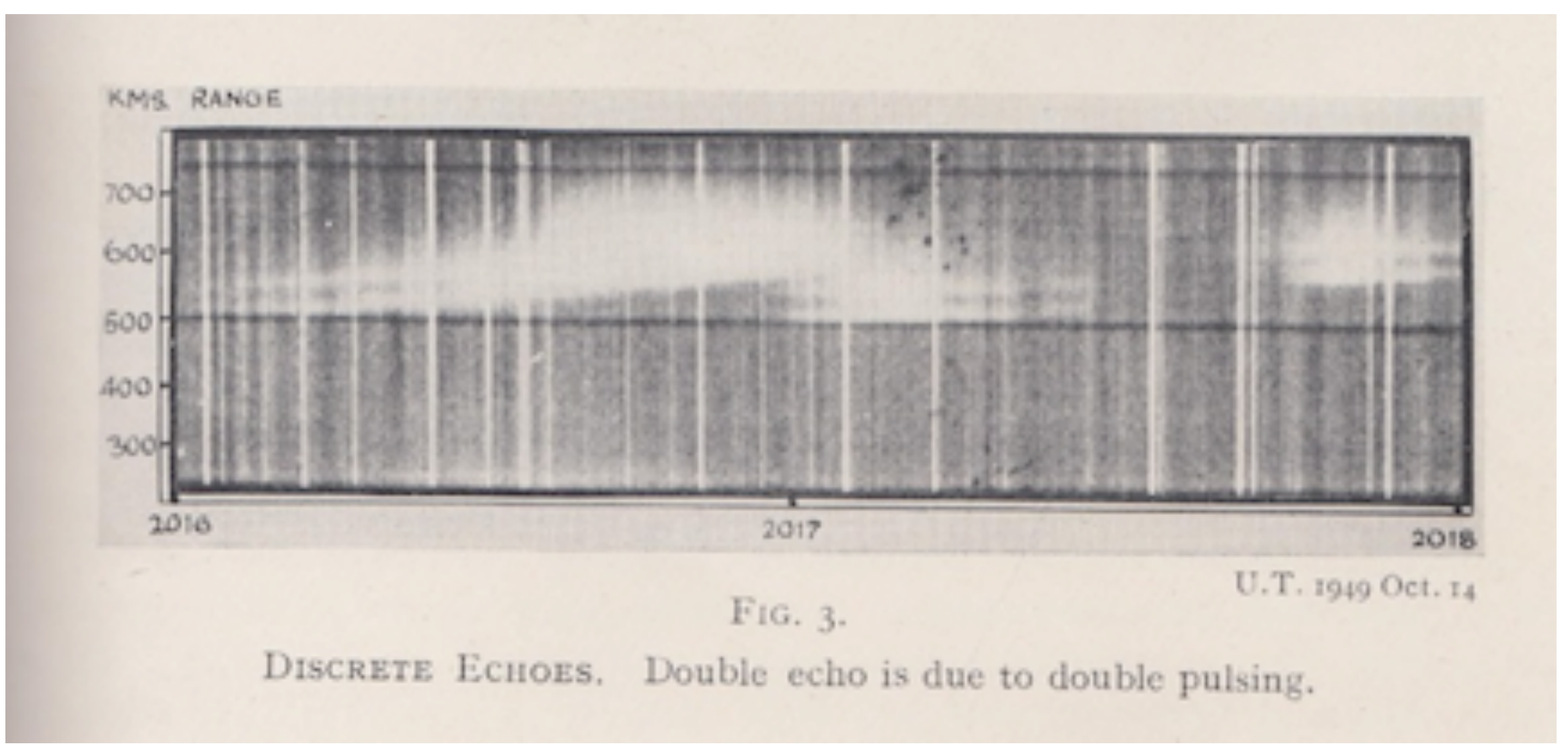

Figure 6: Radar echoes from the aurora of 14 October 1949 obtained by Hawkins. From reference (46)

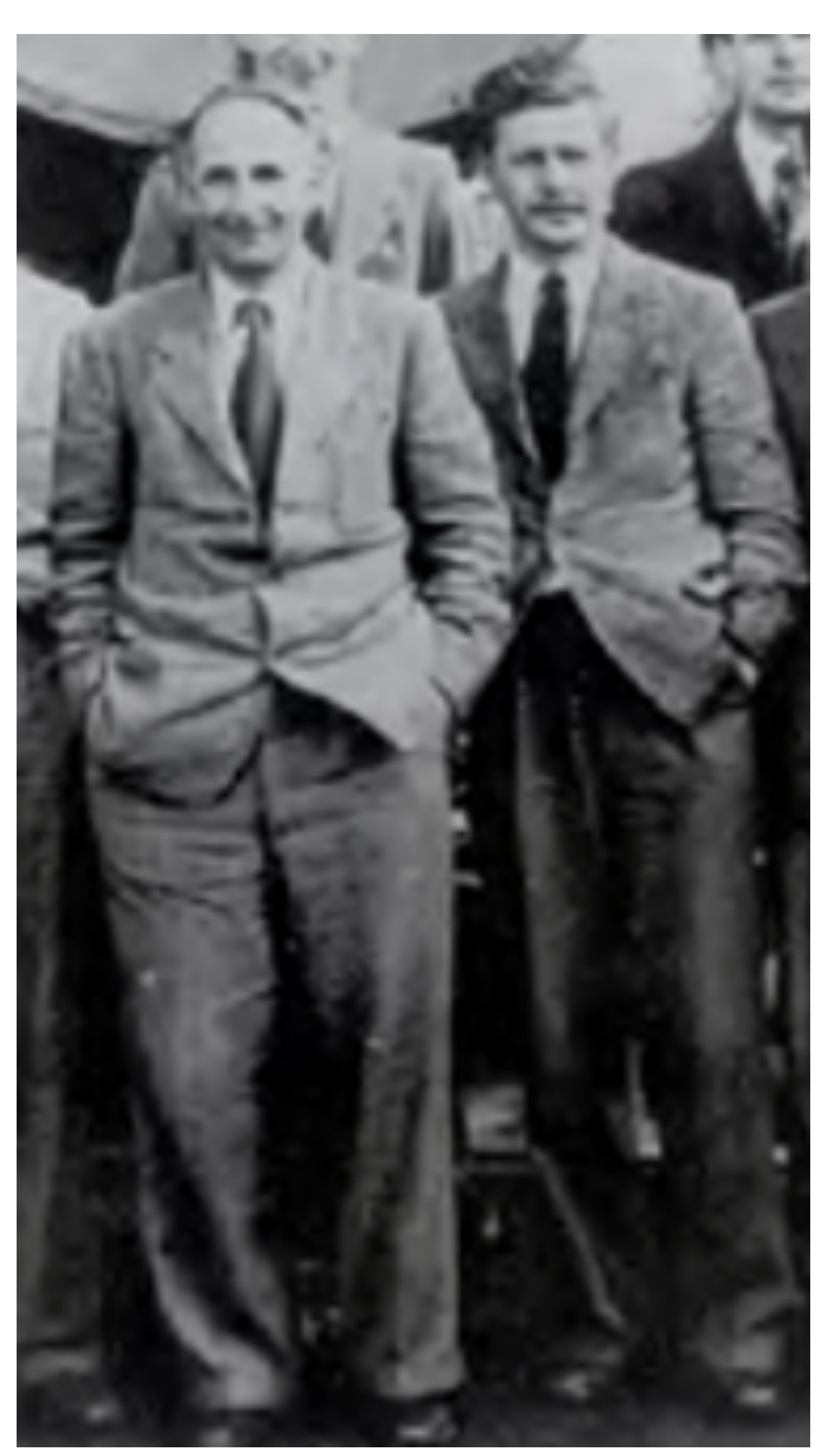

Figure 7: J.A. Clegg with Lovell, 1951. From Figure 1 (the University of Manchester)

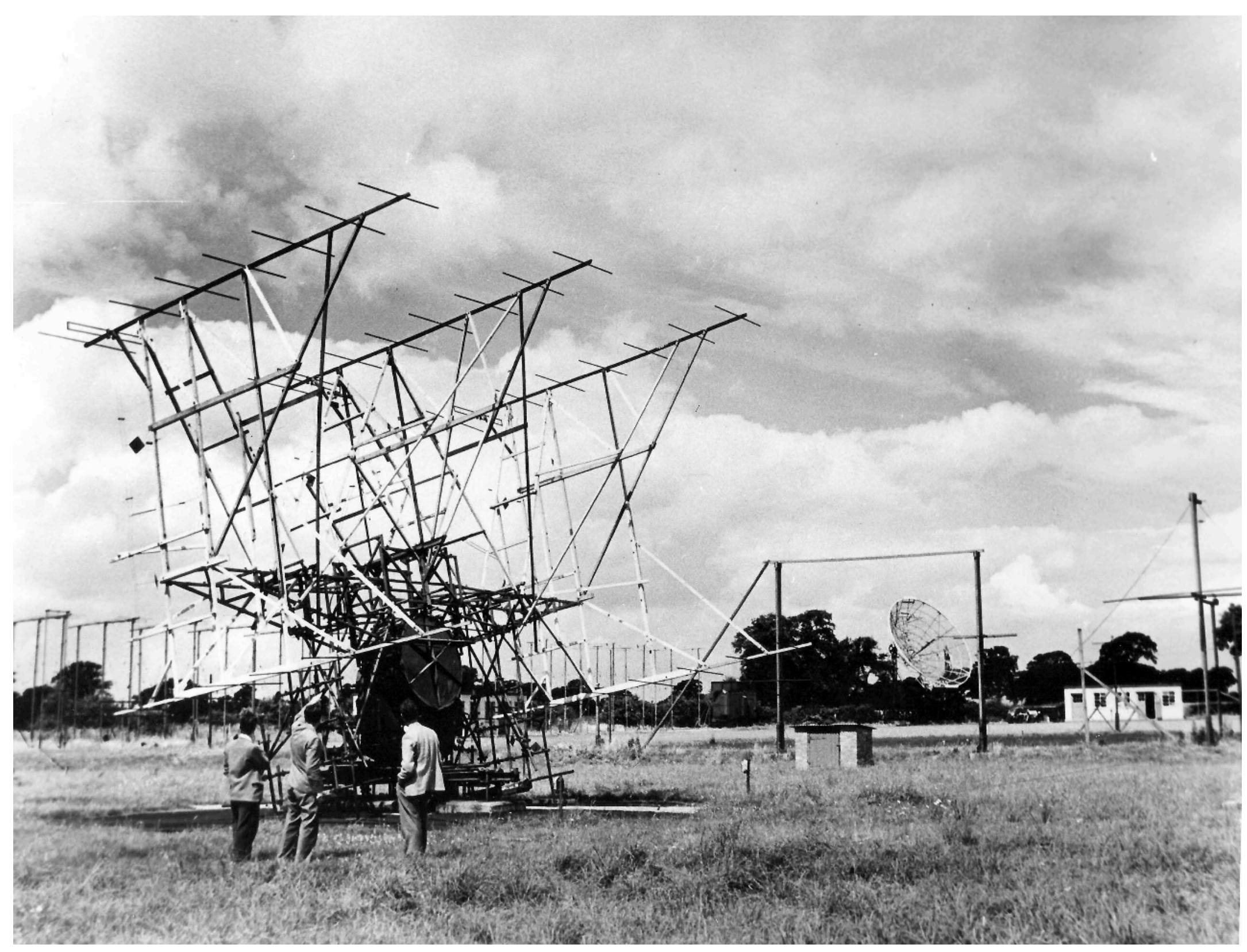

Figure 8: The Searchlight Aerial ca 1950 (the University of Manchester)

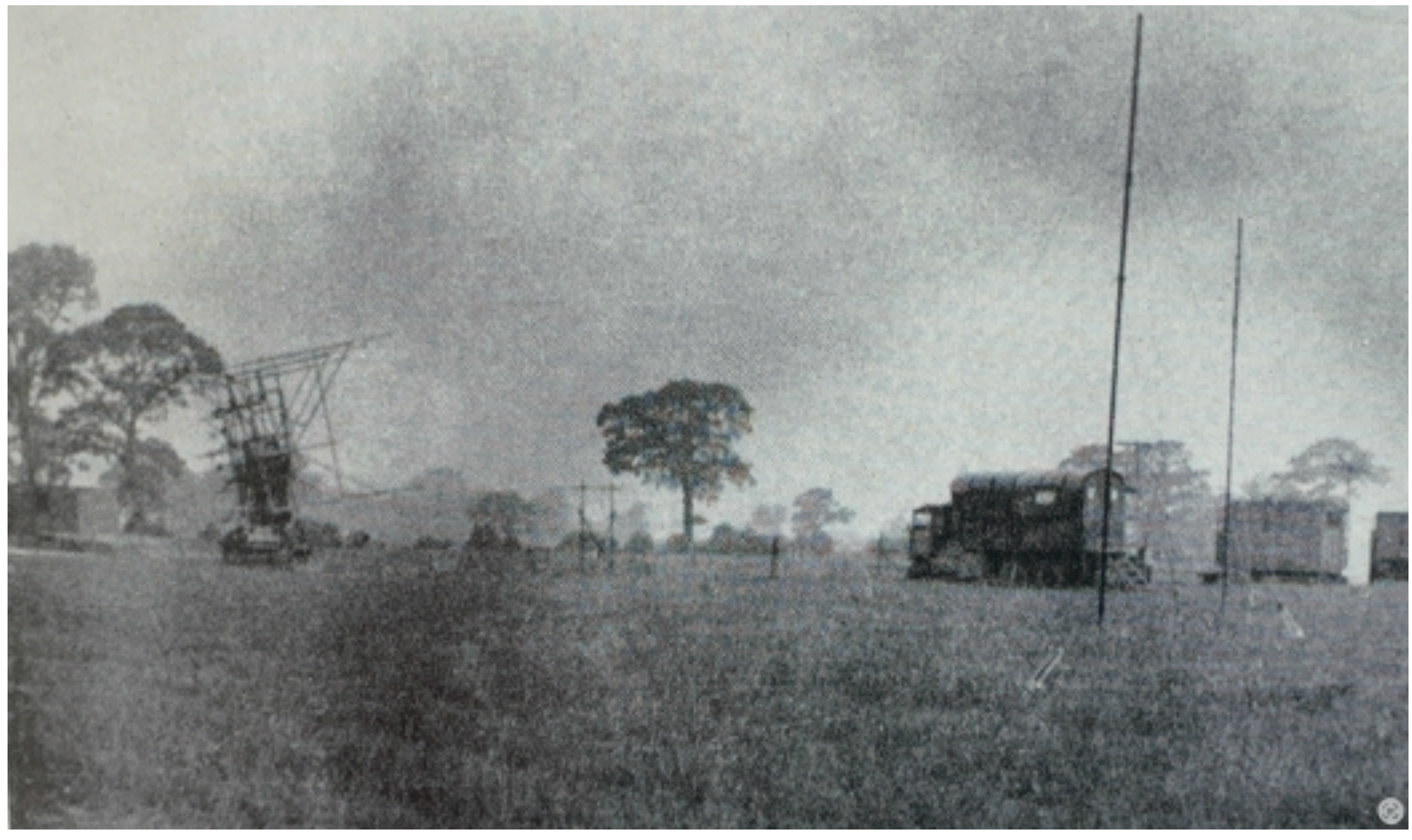

Figure 9: The Searchlight Aerial (left) and the Park Royal (dark truck to right) on the Jodrell Bank observing field, June 1947 (the University of Manchester)

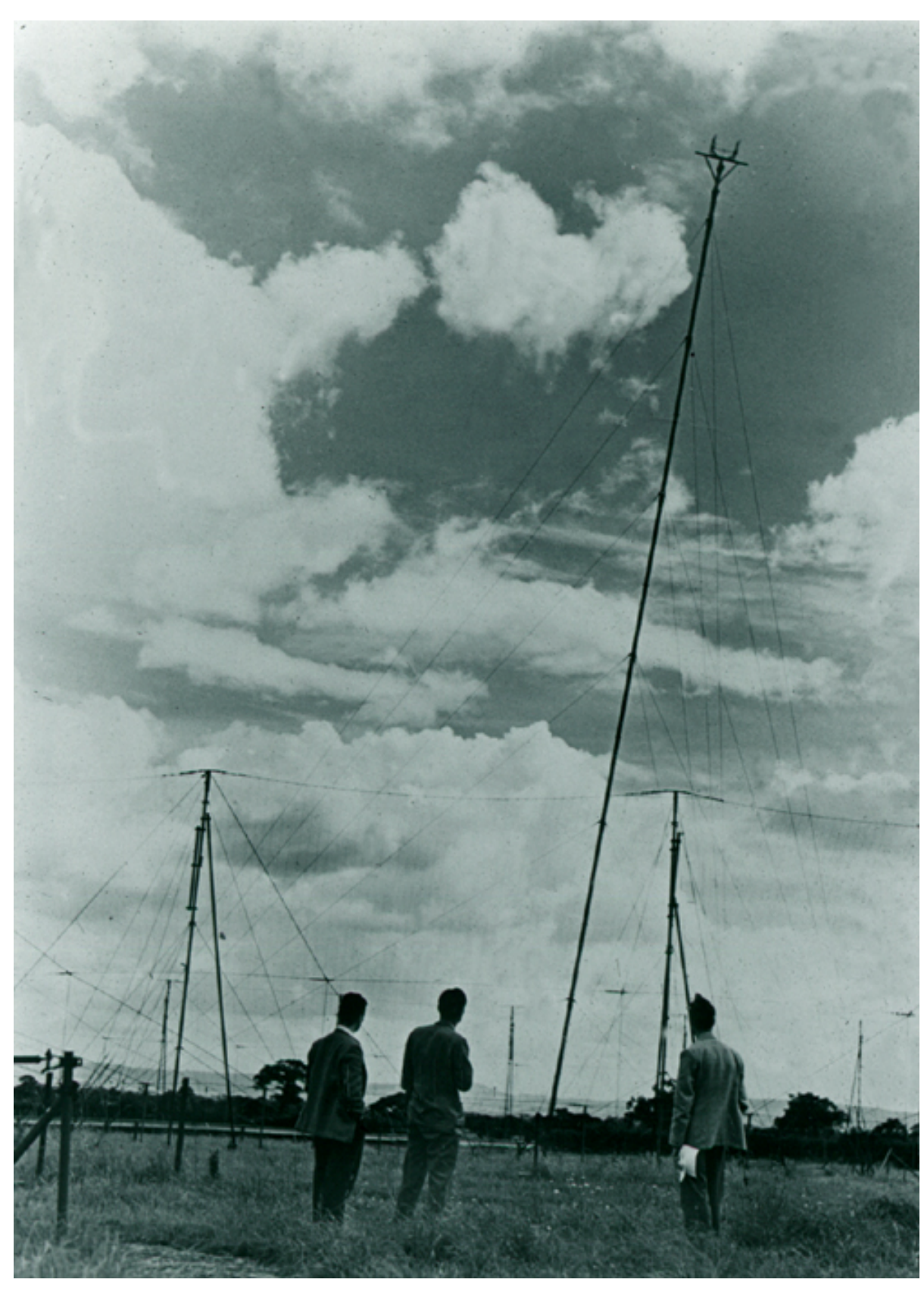

Figure 10: The 218-foot (66 m) Transit Telescope at Jodrell Bank. The 126 ft (38 m) mast is in an almost vertical position. Cyril Hazard is on the left and Robert Hanbury Brown in the centre – both researchers at Jodrell Bank (the University of Manchester)

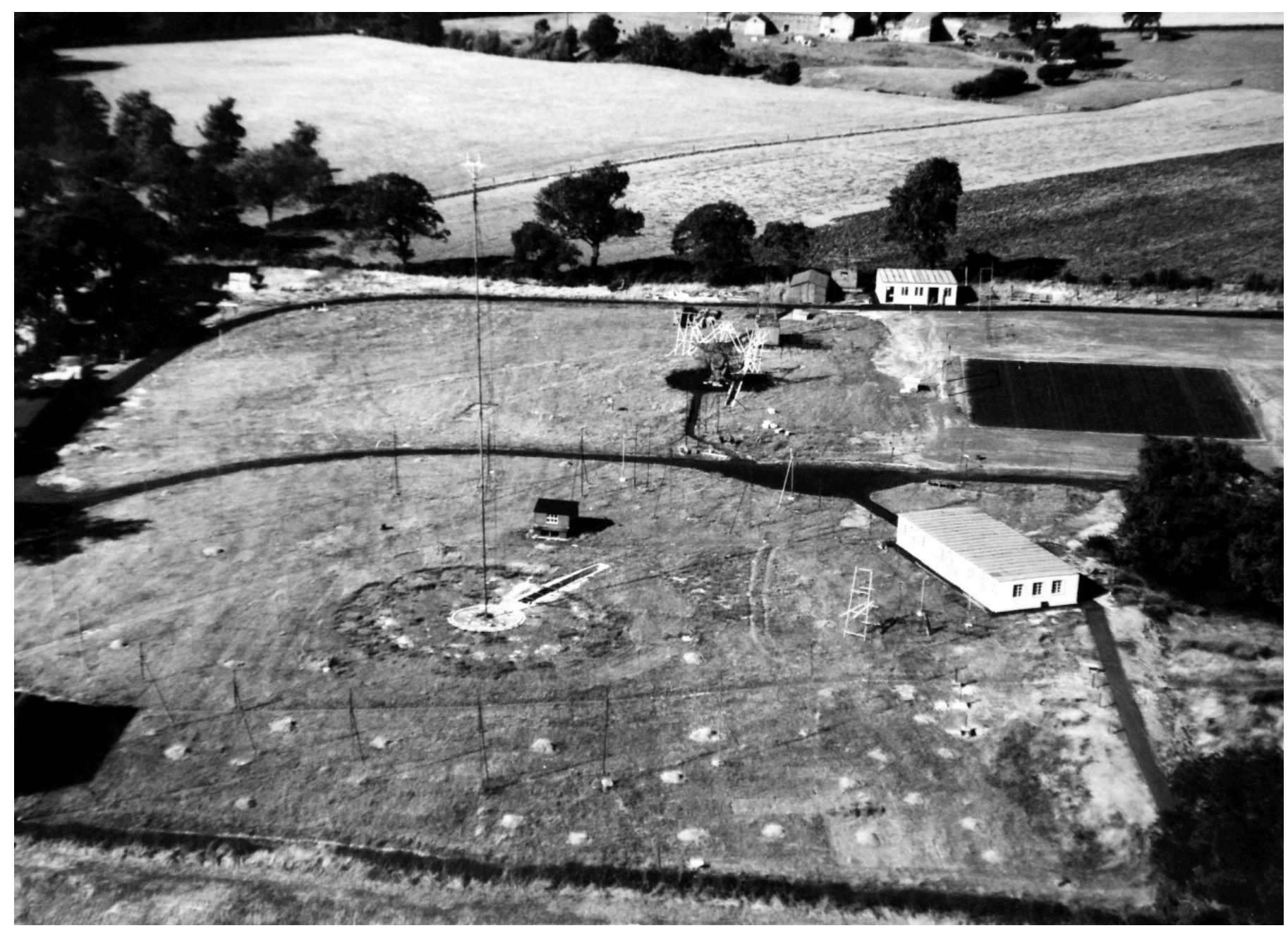

Figure 11: The Jodrell Bank field showing the 218-foot (66 m) Transit Telescope. The Searchlight Aerial is also towards the far side of the field (the University of Manchester)

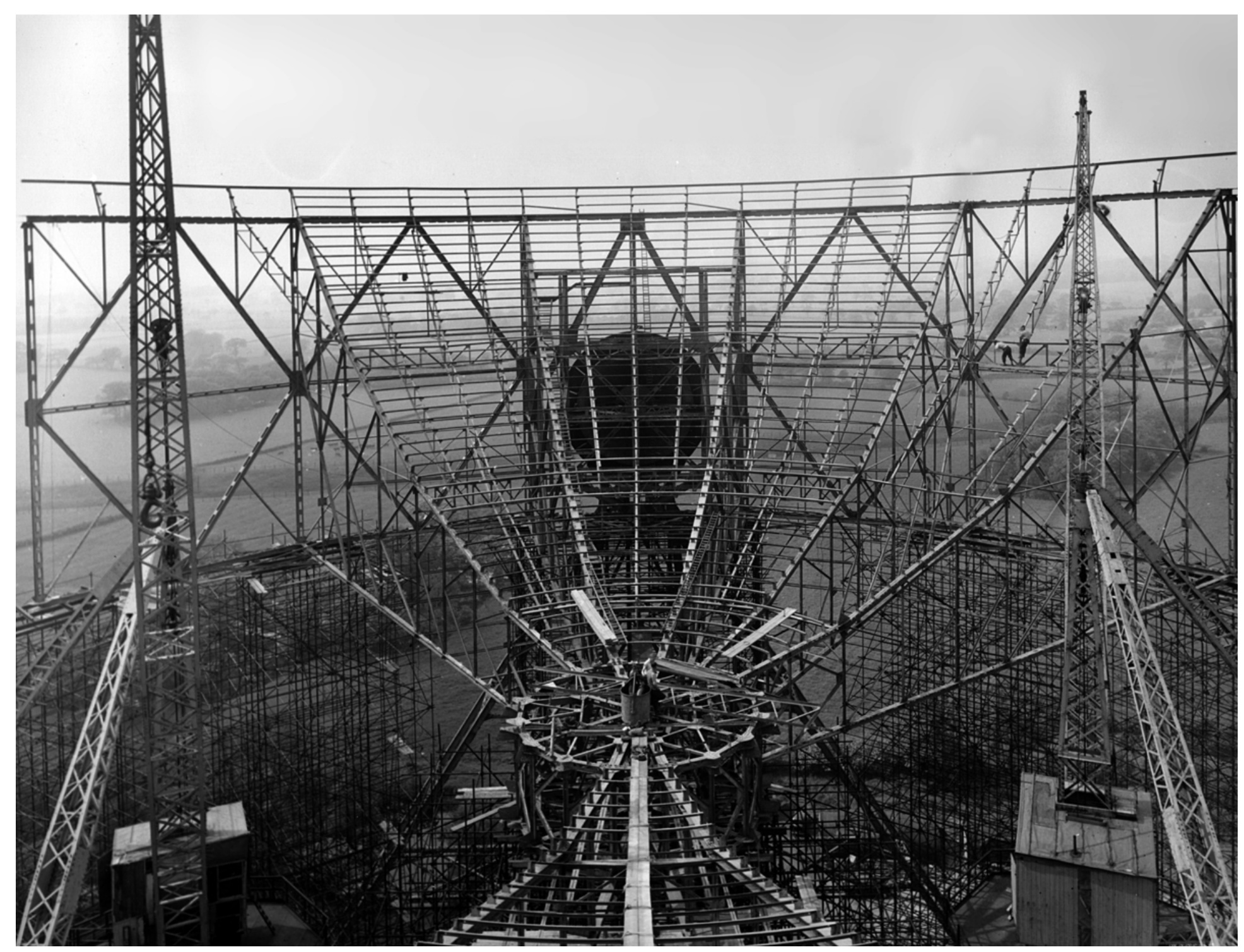

Figure 12: The 250-foot (76 m) Mark 1 telescope rises from the Cheshire Plain, 1956 (the University of Manchester)

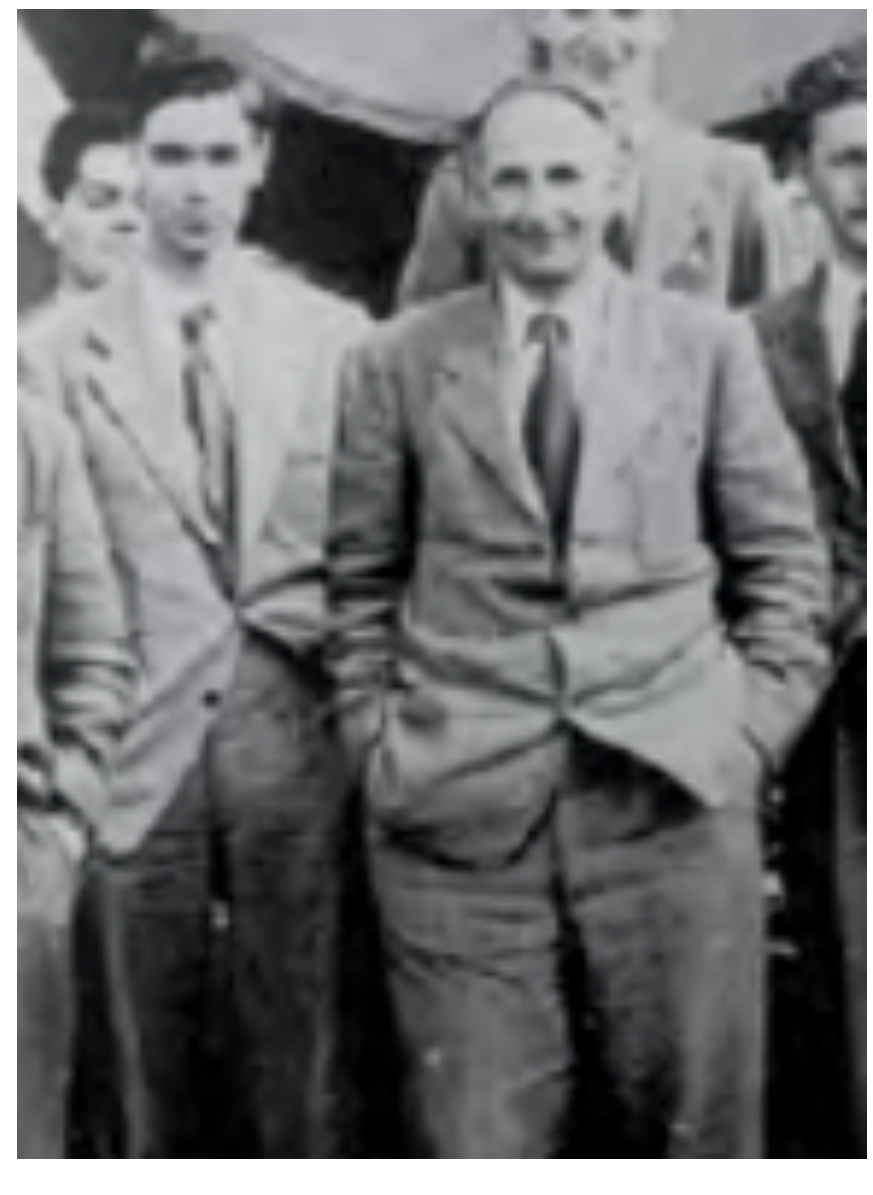

Figure 13: J.G. Davies, left, next to Lovell. From Figure 1 (the University of Manchester)

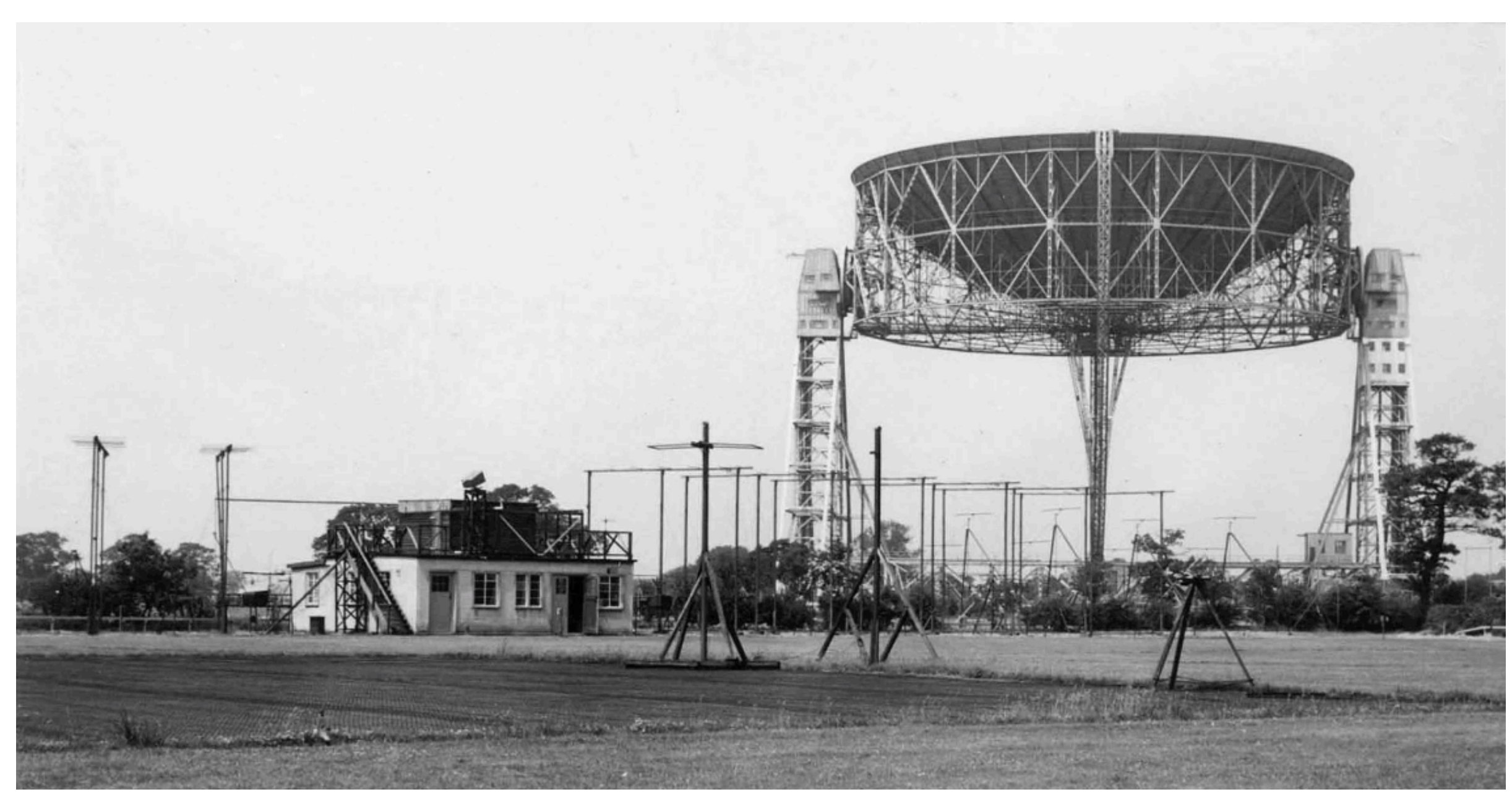

Figure 14: The 250-foot (76 m) *Mark 1* telescope (the University of Manchester)

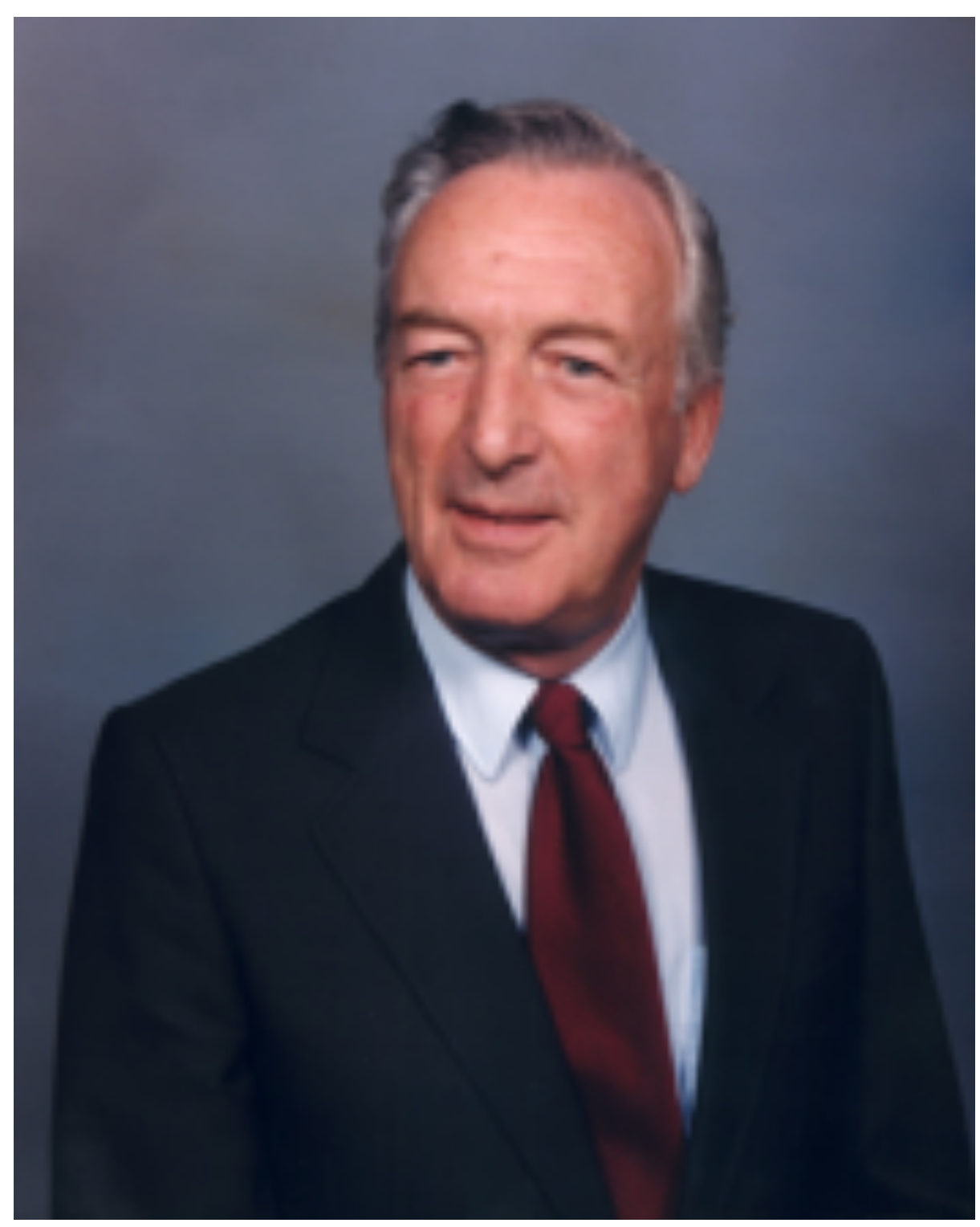

Figure 15: V.A. Hughes (Canadian Astronomical Society)

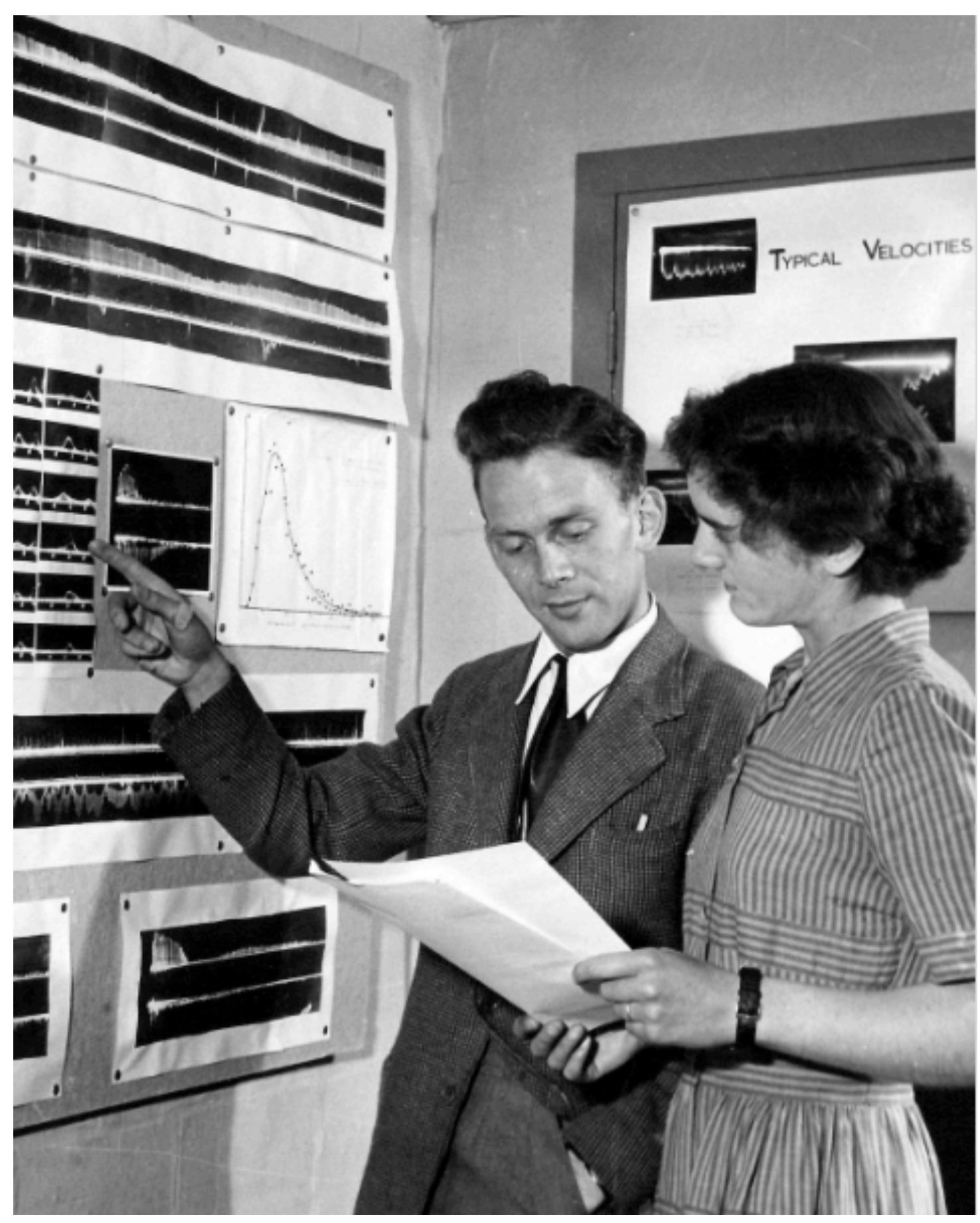

Figure 16: Mary Almond and J.S. (Stanley) Greenhow inside the Radiant Hut at Jodrell Bank, studying film of radar echoes from meteors, 1951 (the University of Manchester)